\documentclass[12pt,draftclsnofoot, onecolumn]{IEEEtran}
\usepackage{amsmath,graphicx}
\usepackage{subfigure}
\usepackage{lineno}
\usepackage{cite}
\usepackage{multirow}
\usepackage{stfloats}
\usepackage{amsfonts}
\usepackage{amssymb}
\usepackage{booktabs} 
\usepackage{makecell}
\usepackage{diagbox}
\usepackage{slashbox}
\usepackage{color}

\usepackage[ruled,linesnumbered,boxed]{algorithm2e}
\begin{document}

\title{A Robust Deep Learning-Based Beamforming Design for RIS-assisted Multiuser MISO Communications with Practical Constraints}

\author{
	\IEEEauthorblockN{
		Wangyang Xu, Lu Gan and 
		Chongwen Huang
		\vspace{-3em}}
	\thanks{The work of Prof. Huang was supported by the China National Key R\&D Program under Grant 2021YFA1000072, National Natural Science Foundation of China under Grant 62101492, Zhejiang Provincial Natural Science Foundation of China under Grant R22F0110230, Zhejiang University Education Foundation Qizhen Scholar Foundation, and Fundamental Research Funds for the Central Universities under Grant 2021FZZX001-21. }
	\thanks{W. Xu  and L. Gan are with the School of Information and Communication Engineering, University of Electronic Science and Technology of China, Chengdu, Sichuan, 611731, China (e-mail: wangyangxu@std.uestc.edu.cn;ganlu@uestc.edu.cn).}
	\thanks{C. Huang is with College of Information Science and Electronic Engineering, Zhejiang University, Hangzhou 310027, China, and with International Joint Innovation Center, Zhejiang University, Haining 314400, China, and also with Zhejiang Provincial Key Laboratory of Info. Proc., Commun. \& Netw. (IPCAN), Hangzhou 310027, China. (e-mail: chongwenhuang@ieee.org).}
}

\maketitle

\begin{abstract}
Reconfigurable intelligent surface (RIS) has become a promising technology to improve wireless communication in recent years. It steers the incident signals to create a favorable propagation environment by controlling the reconfigurable passive elements with less hardware cost and lower power consumption. In this paper, we consider a RIS-aided multiuser multiple-input single-output downlink communication system. We aim to maximize the weighted sum-rate of all users by joint optimizing the active beamforming at the access point and the passive beamforming vector of the RIS elements. Unlike most existing works, we consider the more practical situation with the discrete phase shifts and imperfect channel state information (CSI). Specifically, for the situation that the discrete phase shifts and perfect CSI are considered, we first develop a deep quantization neural network (DQNN) to simultaneously design the active and passive beamforming while most reported works design them alternatively. Then, we propose an improved structure (I-DQNN) based on DQNN to simplify the parameters decision process when the control bits of each RIS element are greater than 1 bit. Finally, we extend the two proposed DQNN-based algorithms to the case that the discrete phase shifts and imperfect CSI are considered simultaneously. Our simulation results show that the two DQNN-based algorithms have better performance than traditional algorithms in the perfect CSI case, and are also more robust in the imperfect CSI case.  
\end{abstract}

\begin{IEEEkeywords}
Reconfigurable intelligent surface, joint beamforming, deep learning, quantization layer, discrete phase shifts, imperfect CSI.
\end{IEEEkeywords}

%
\IEEEpeerreviewmaketitle

\section{Introduction}
In recent years, with the explosion of mobile devices in wireless communications, it has become increasingly important to improve communication quality of service (QoS) and communication speed. Therefore, numerous works have been proposed to meet these requirements, such as the relaying \cite{song2011relay} and massive multiple-input and multiple-output (MIMO) systems \cite{larsson2014massive}. The former actively enhances the signal strength but needs extra hardware consumption and complex signal processing. The latter uses a large number of antennas to exploit the spatial gain, which can effectively improve the spectrum efficiency. However, implementing massive MIMO is still a very challenging task since the high cost, increased power consumption, and constrained physical size and shape. 

Reconfigurable intelligent surface (RIS) is a promising technology that greatly improves coverage, throughput, spectrum efficiency, and energy efficiency of wireless communication \cite{1WQQtoward2020,wu2019beamforming,Anew,WCT,huang2020holographic}.  The RIS does not require active radio frequency (RF) chains or complex signal processing. Instead, it relies on passive components with simpler hardware and less energy consumption than traditional transceivers and relay systems to reflect signals. These passive elements can be controlled by simple programmable positive-intrinsic-negative (PIN) diodes and a micro-controller \cite{cui2014coding}. The micro-controller adjusts the phase shift and amplitude of each passive element at the RIS to achieve fine-grained three-dimensional passive beamforming so that the signal power is improved at the desired receiver or reduced at the non-desired receiver \cite{nadeem2019intelligent}.

Although the RIS has so many excellent performances and wide application prospects in future wireless communications, it still faces many daunting challenges. One of them is the joint designing of the active beamforming at the AP and the passive beamforming at the RIS in the RIS-aided wireless communication system with practical constraints.  

Recently, there are plenty of works that have studied the passive beamforming issues from different aspects and setups. In \cite{wu2019beamforming,guo2020weighted,10MISOCEBO,12xu}, the authors considered that each element at the RIS has continuous phase shifts for the passive beamforming. Specifically, \cite{wu2019beamforming} solved the joint beamforming problem by decomposing the non-convex coupling objective into active transmit beamforming and passive beamforming problems, respectively. For active transmit beamforming, the maximum ratio transmission (MRT) and the minimum mean squared error (MMSE) criterion were used in the single-user and the multi-user situations.  For passive beamforming, the semidefinite relaxation (SDR) method was applied. \cite{guo2020weighted} maximized the weighted sum-rate (WSR) of all users by joint designing the active beamforming and the passive beamforming. \cite{10MISOCEBO} proposed an alternating optimization (AO) algorithm to maximize the minimum signal to interference plus noise ratio (SINR), subject to a transmit power constraint and a unit-modulus constraint. \cite{12xu} proposed two computationally affordable approaches for passive beamforming, capitalizing on alternating maximization, gradient descent search, and sequential fractional programming. \cite{Zheng2020Intelligent} considered a continuous phase-shift passive beamforming problem in an orthogonal frequency division multiplexing system and the reflection coefficients are then optimized by a low-complexity algorithm based on the resolved strongest signal path in the time domain.  

However, it is impractical to achieve continuous phase shifts due to the hardware limitation. Therefore, there are many works that have taken into account the discrete phase-shift situation, such as  \cite{13xu,14xu,15xu,di2020hybrid}. Specifically, \cite{13xu} relaxed the discrete phase-shift constraint to the continuous phase-shift constraint. Then, the final results were obtained by discretizing the continuous phases. In \cite{14xu}, the non-convex WSR problem with discrete phase shifts was firstly decoupled via Lagrangian dual transform, and then the active and passive beamforming can be optimized alternatingly. \cite{15xu} formulated the problem as minimizing the transmit power, and optimized the continuous transmit precoding at the AP and the discrete phase shifts at the RIS. \cite{di2020hybrid} proposed a hybrid beamforming scheme and formulated a sum-rate maximization problem with a limited number of discrete phase shifts. An iterative algorithm was designed to solve this problem.

Most existing works assumed that the CSI between AP/BS and RIS and RIS and users is perfect \cite{wu2019beamforming,12xu,16xu,perfect1,perfect2,perfect3,an1,an2}. However, this assumption is impractical because of the complexity of channel estimation. Furthermore, there have been some works on joint beamforming in the imperfect CSI case. In \cite{guo2020weighted}, the joint beamforming problem with imperfect CSI was formulated as maximizing the expectation of the achievable WSR. The sample average approximation method and the stochastic successive convex approximation (SCA) technique were applied to solve this problem.  In \cite{8mainYou}, the authors applied the least square method for channel estimation, explicitly reflecting the channel estimation's error in the SINR expression, which was solved by the SDR-based method with high complexity. In \cite{revised2}, the authors exploited some algorithms such as alternating optimization, a penalty-based approach, SCA, and semidefinite relaxation to design the secure wireless communications assisted by multiple IRSs under the imperfect CSI constraint. In \cite{revised1}, the secure communication system with self-sustainable IRS was considered under the discrete phase shift and imperfect CSI constraints. The complex optimization objective was solved by decoupling the coupled variables.

To optimize active beamforming and passive beamforming simultaneously with less complexity, deep learning (DL) technologies have been applied for RIS-aided communication systems in recent years. For example, \cite{7taha2019enabling} proposed a supervised learning (SL) method with four layers for the passive beamforming design, which reduces the pilot training overhead but requires active RIS elements for channel training, while the authors reduced complexity at the model training stage via unsupervised learning in \cite{17xu}. Moreover, \cite{18xu} introduced a deep Q network (DQN) which does not require labels like SL but with longer training time. Note that \cite{18xu} also needs active RIS elements for channel acquisition like \cite{7taha2019enabling}. Some deep deterministic policy gradient based algorithms have been proposed in \cite{huang2020reconfigurable,19xu,20xu}. They all have better performance than DQN-based methods. Besides, the DL-based methods are also applied in many more complex scenarios. For example, \cite{DL14} proposed a DQN with three layers to beat eavesdropping robustly but with high model training complexity in secure beamforming. \cite{DL3} focused on energy-efficient beamforming with DQN, and the indoor beamforming was considered in \cite{DL15} via a DL-based method.

Currently, there are many DL-based methods for the optimization of continuous RIS phase shifts with the perfect CSI, but as far as we know, there is no DL-based method for the optimization in a practical situation that the discrete phase shifts and imperfect CSI are considered yet.

Motivate by these, in this paper, we consider the joint beamforming problem with discrete phase shifts and the imperfect CSI in a RIS-aided multiuser multiple-input single-output (MISO) downlink communication system. We develop a deep quantization neural network (DQNN) and its improved structure (I-DQNN) to obtain a near-optimal solution of the WSR problem with discrete phase shifts in both perfect CSI and imperfect CSI cases. Finally, the performance of DQNN and I-DQNN are evaluated in different setups. The main contributions of this paper are summarized as follows:

\begin{itemize}
	\item Firstly, we formulate the joint active and passive beamforming problem as a practical WSR problem with the discrete phase shifts and perfect CSI in a multiuser MISO downlink communication system. Then, we propose a DQNN-based algorithm to solve the WSR problem. The DQNN can simultaneously predict active and passive beamforming instead of designing them alternatively like existing AO-based algorithms, which might be trapped in weak stationary points over iterations. We also propose a scalar quantization layer, which can be applied directly in the stochastic gradient descent-based DQNN to take the errors caused by a quantization into joint optimization, thereby improving the performance of RIS-assisted systems.
	\item Secondly, we analyze the mismatch problem of the DQNN-based algorithm and solve this problem to obtain optimal parameters of DQNN via a comparative search method. Besides, we also develop an improved DQNN (I-DQNN) by applying a penalty term to lower the overhead of the parameters' optimization for more efficient training in the face of 1-bit quantized discrete phase shift. Simulation results show the efficiency of the proposed I-DQNN.
	\item Finally, for the discrete phase shifts and the imperfect CSI case, we formulate the joint beamforming problem with an expectation operator as a more straightforward WSR problem, which our proposed DL-based algorithms can readily solve. Simulation results show that the two algorithms outperform traditional methods, albeit their significantly reduced computational complexity.
\end{itemize}

The rest of this paper is organized as follows. Section II introduces the system model and the problem formulation for the RIS-aided multiuser MISO downlink communication system. In Sections III, we introduce the structure and application of DQNN and I-DQNN. Section IV presents the simulation results of our proposed algorithms. Finally, we conclude the paper in Section V.

\emph{Notations:} Scalars are denoted by italic letters. Vectors and matrices are denoted by bold-face lower-case and upper-case letters, respectively. The superscripts ${\left(  \cdot  \right)^T}$ and ${\left(  \cdot  \right)^H}$ denote the operations of transpose and Hermitian transpose. $\mathbb{E}{\left[  \cdot  \right]}$ denotes the statistical expectation. $\left|  \cdot  \right|$ denotes the absolute value of a real number. $\left\|  \cdot  \right\|$ denotes the 2-Norm of a vector or a matrix. For a complex number, $\operatorname{Re} \left\{  \cdot  \right\}$ and $\operatorname{Im}  \left\{  \cdot  \right\}$ denote the real part and imaginary part, and $\angle \left(  \cdot  \right)$ denotes the phase of it. Besides, for a matrix, $diag\left(  \cdot  \right)$ denotes a square matrix, and the elements on its main diagonal are consisted of the elements in $\left(  \cdot  \right)$. Besides, ${\left[  \cdot  \right]_{m,n}}$ denotes the element at the $m$-th row and the $n$-th column of a matrix. ${\left(  \cdot  \right)^{'}}$ represents the derivative of a function, and $f\left(  \cdot  \right)$ denotes a function.

\section{System model and problem formulation}

\subsection{System Model}
\begin{figure}[tbp]
	\centering {
		\begin{tabular}{ccc}
			\includegraphics[width=0.6\textwidth]{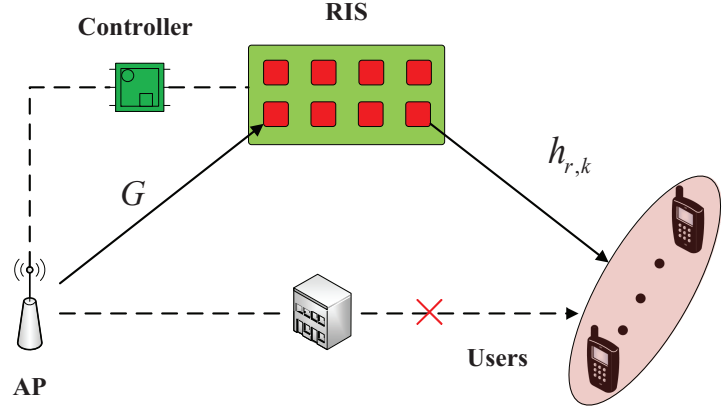}\\
		\end{tabular}
	}
	\caption{A RIS-assisted MISO communication system.}
	\vspace{-0.45\baselineskip}
	\label{fig1}
\end{figure}

In this paper, we consider a basic RIS-aided multiuser MISO system as illustrated in Fig. \ref{fig1}. The system consists of one AP with $M$ antennas, one micro-controller, $K$ single-antenna users and one RIS. The RIS is equipped with $N = {N_x} * {N_y}$ passive reflecting elements in a rectangular arrangement. $N_x$ and $N_y$ represent the number of elements in each row and each column, respectively. During the downlink period, the $M$ antennas of the AP transmit $K$ data streams to different $K$ users. We assume that the direct links between the AP and the users are blocked by some obstacles, which is practical in many complex environments like indoor. Note that our work is also adapted to the situation where direct links exist by simply adjusting the loss function and network input.

Let ${\boldsymbol{G}} \in {\mathbb{C}^{N \times M}}$ and ${\boldsymbol{h}_{r,k}} \in {\mathbb{C}^{N \times 1}}, k = 1,...,K$ denote the channel matrices between the AP and the RIS, and the RIS and the $k$-th user, respectively. For simplicity, we assume that all the channels experience quasistatic flat-fading and the CSI is known at both AP and RIS \cite{wei2020channel}.   Let ${\boldsymbol{\theta} ^H} \buildrel \Delta \over = [{v_1}{e^{j{\varphi_1}}},{v_2}{e^{j{\varphi_2}}}, \cdots ,{v_N}{e^{j{\varphi_N}}}]$ denotes the reflection coefficient vector of the reflecting elements at the RIS where ${v_l}$ and ${{\varphi_l}}$ are the amplitude and phase shift, respectively.  For simplicity, we assume that the RIS can completely reflect the incident signals, which means that the amplitudes of all reflection coefficients are constant and set as ${v_l} = 1,\;l = 1,2, \ldots ,N$. Thus, the reflection coefficient vector can be rewritten as ${\boldsymbol{\theta} ^H} \buildrel \Delta \over = [{e^{j{\varphi_1}}},{e^{j{\varphi_2}}}, \cdots ,{e^{j{\varphi_N}}}]$. In practice, only a finite number of discrete values for phase shifts are available. We assume that $b$ denotes the bits controlling per element and $B = {2^b}$ is the discrete phase-shift levels. Then we can uniformly quantize the continuous phase shifts in the range of $[0,2\pi )$, and the set of these phase shifts is $S \buildrel \Delta \over = \{ 0,\Delta w, \cdots ,(B - 1)\Delta w\}$, where $\Delta w = \frac{{2\pi }}{B}$.

The signal received at user $k$ in a downlink transmission can be expressed as 
\begin{equation}   \label{xu1}
\left.{y_k} = \boldsymbol{h}_{r,k}^H\boldsymbol{\Theta} {\boldsymbol{G}}{\boldsymbol{W}}{\boldsymbol{x}} + {z_k},\;k \in \{ 1,2, \cdots ,\;K\},\right.
\end{equation}
where $\boldsymbol{\Theta} = diag({e^{j{\varphi _1}}},{e^{j{\varphi_2}}}, \cdots ,{e^{j{\varphi_N}}})$ is the phase shift matrix of the RIS,  ${\boldsymbol{W}} \in {\mathbb{C}^{M \times K}}$ is the active transmit beamforming at the AP,  ${\boldsymbol{x}} \in {\mathbb{C}^{K \times 1}}$ is the vector of the transmitted data streams, with zero mean unit variance entries, $\mathbb{E}\{ {\left| {{\boldsymbol{x}}} \right|^2}\}  = 1$, and ${z_k} \sim {\cal C}{\cal N}(0,\sigma _k^2)$ is the additive white
Gaussian noise (AWGN) in the receiver of $k$-th user. For simplicity, we assume that $\sigma _k^2 = \sigma ^2$.

The received signal model (\ref{xu1}) can be further written as:
 \begin{equation}   \label{xu2}
 \left.{y_k} = \boldsymbol{h}_{r,k}^H\boldsymbol{\Theta} {\boldsymbol{G}}{\boldsymbol{w}_k}{x_k} + \sum\limits_{n,n \ne k}^K {\boldsymbol{h}_{r,k}^H\boldsymbol{\Theta} {\boldsymbol{G}}{\boldsymbol{w}_n}{x_n}}  + {z_k},\right.
 \end{equation}
 where ${\boldsymbol{w}_k}$ is the $k$-th column vector of the active transmit beamforming matrix $\boldsymbol{W}$. In (\ref{xu2}), $\boldsymbol{h}_{r,k}^H\boldsymbol{\Theta} {\boldsymbol{G}}{\boldsymbol{w}_k}{x_k}$ is the desired signal at the $k$-th user and $\sum\limits_{n,n \ne k}^K {\boldsymbol{h}_{r,k}^H\boldsymbol{\Theta} {\boldsymbol{G}}{\boldsymbol{w}_n}{x_n}}$ is treated as cochannel interference.
 
 Then the SINR at the $k$-th user is given by 
 \begin{equation}   \label{xu3}
 \left.{\gamma _k} = \frac{{{{\left| {\boldsymbol{h}_{r,k}^H\boldsymbol{\Theta} {\boldsymbol{G}}{\boldsymbol{w}_k}{x_k}} \right|}^2}}}{{\sum\limits_{n,n \ne k}^K {{{\left| {\boldsymbol{h}_{r,k}^H\boldsymbol{\Theta} {\boldsymbol{G}}{\boldsymbol{w}_n}{x_n}} \right|}^2} + {\sigma ^2}} }}.\right.
 \end{equation} 
 
 Besides, the transmit power constraint of AP is
  \begin{equation}   \label{xu4}
 \left.\sum\limits_{k = 1}^K {{{\left\| {{\boldsymbol{w}_k}} \right\|}^2}}  \le {P_t}\right..
 \end{equation}
 
 \subsection{Problem Formulation}
 All works we have done for the joint beamforming problem in this paper are based on maximizing the WSR of all the users, subject to the transmit power and discrete phase shifts constraints. At first, we consider this problem under the assumption that the CSI is perfect at both AP and RIS. Then we extend it to the more practical and complex case that only imperfect CSI is available at both AP and RIS.
  \subsubsection{Perfect CSI}
 Although it is impractical to consider perfect CSI, it helps measure the proposed algorithm's fundamental performance and be used as a benchmark method when considering the joint beamforming problem. More importantly, many SL-based methods use beamforming vectors obtained in perfect CSI as labels for training. Therefore, we can also provide labels for SL in this way \cite{liaskos2019interpretable,jia2020machine}.
 
 We extend the WSR maximization problem proposed in \cite{guo2020weighted} to a new one with the discrete phase shifts constraint, which is given by 
 \begin{equation}   \label{xu5}
 \left.\begin{array}{l}
 {\rm{P}}1:\;\;\;\;\mathop {\max }\limits_{\boldsymbol{W},\boldsymbol{\theta} } \;\;\;\;  {f_A}\left( {\boldsymbol{W},\boldsymbol{\theta} } \right) = \sum\limits_{k = 1}^K {{q _k}} \log_2 \left( {1 + {\gamma _k}} \right)\\
 \;\;\;\;\;\;\;\;\;\;s.t.\;\;\;\;\left| {{{\theta} _n}} \right| = 1,\;\;\;\;\;\;\;{\forall n} = 1,2, \cdots ,N,\\
 \;\;\;\;\;\;\;\;\;\;\;\;\;\;\;\;\;\;\;\angle {{\theta} _n} \in S,\;\;\;\;\;\;\;{\forall n} = 1,2, \cdots ,N,\\
 \;\;\;\;\;\;\;\;\;\;\;\;\;\;\;\;\;\;\;\sum\limits_{k = 1}^K {{{\left\| {{\boldsymbol{w}_k}} \right\|}^2}}  \le {P_t}.
 \end{array}\right.
 \end{equation}
 where ${f_A}\left( {\boldsymbol{W},\boldsymbol{\theta} } \right)$ is the WSR of all users, ${q _k}$ is used to represent the priority of the $k$-th user and ${{\theta} _n}$ is the $n$-th reflection coefficient of the RIS.
 It can be seen that  (\ref{xu5}) is a non-convex optimization problem due to the non-convex objective function and the constraints. The optimal result can be obtained by the exhaustive search method. However, it is nearly impossible when the number of elements of RIS is very large. Therefore, many existing works try to find the suboptimal solution based on the AO algorithms. Specifically, in each iteration, the AO algorithms optimize the $\boldsymbol{W}$ with fixed $\boldsymbol{\theta}$ and then optimize the $\boldsymbol{\theta}$ with the fixed $\boldsymbol{W}$. The difference between them is on the aspects of methods when optimizing $\boldsymbol{W}$ and  $\boldsymbol{\theta}$. Except for the max-sum-rate problem considered in this paper, P1 can also be extended to solve the problem having minimum individual user data rate constraints. For example, minimum user data rate constraints can be easily transformed and added to the objective optimization function and solved by the proposed deep learning method.
  \subsubsection{Imperfect CSI}
 To optimize the passive beamforming vector $\boldsymbol{\theta}$, the channels should be estimated firstly. However, it is difficult to obtain an accurate CSI due to the complexity of channel estimation. Therefore, the imperfect CSI case is more practical. In this paper, we assume that the CSI obtained at both AP and RIS is imperfect. Then, the channel estimation errors ${\boldsymbol{h}_e}$ and ${\boldsymbol{G}_e}$ can be expressed as 
  \begin{equation}   \label{xu6}
 \left.\begin{array}{l}
 {\boldsymbol{h}_e} = {\boldsymbol{h}_{r,k}} - {{\hat {\boldsymbol{h}}}_{r,k}},\\
 {\boldsymbol{G}_e} = \boldsymbol{G} - \hat {\boldsymbol{G}},
 \end{array}\right.
 \end{equation}
 where $\boldsymbol{h}_{r,k}$ and $\boldsymbol{G}$ are the true channels, ${{\hat {\boldsymbol{h}}}_{r,k}}$ and $\hat {\boldsymbol{G}}$ are the estimated channels. Note that the specific channel estimation mechanism is not discussed since it is not our paper's focus. We just assume that the MMSE method is applied in the channel estimation, which makes the estimation errors and the estimated channels irrelevant \cite{wang2006adaptive,dabbagh2008multiple}. 
 
 Then, (\ref{xu2}) can be further expressed as
 \begin{align}   \label{xu7}
 {y_k} &= ({\hat {\boldsymbol{h}}_{r,k}^H + {\boldsymbol{h}_e}^H})\boldsymbol{\Theta} ({\hat {\boldsymbol{G}}} + {\boldsymbol{G}_e}){\boldsymbol{w}_k}{x_k}\notag\\
 &+ \sum\limits_{n,n \ne k}^K {({\hat {\boldsymbol{h}}_{r,k}^H + {\boldsymbol{h}_e}^H})\boldsymbol{\Theta} {({\hat {\boldsymbol{G}}} + {\boldsymbol{G}_e})}{\boldsymbol{w}_n}{x_n}}  + {z_k}.
 \end{align}

Similarly, (\ref{xu3}) is rewritten as 
 \begin{equation}   \label{xu8}
\left.{\gamma _k} = \frac{{{{\left| {({\hat {\boldsymbol{h}}_{r,k}^H + {\boldsymbol{h}_e}^H})\boldsymbol{\Theta} {({\hat {\boldsymbol{G}}} + {\boldsymbol{G}_e})}{\boldsymbol{w}_k}{x_k}} \right|}^2}}}{{\sum\limits_{n,n \ne k}^K {{{\left| {({\hat {\boldsymbol{h}}_{r,k}^H + {\boldsymbol{h}_e}^H})\boldsymbol{\Theta} {({\hat {\boldsymbol{G}}} + {\boldsymbol{G}_e})}{\boldsymbol{w}_n}{x_n}} \right|}^2} + {\sigma ^2}} }}.\right.
\end{equation} 

  We assume that the estimation errors ${\boldsymbol{h}_e}$ and ${\boldsymbol{G}_e}$ follow the circularly symmetric complex gaussian (CSCG) distribution. Therefore, the true channels ${\boldsymbol{h}}_{r,k}$ and ${\boldsymbol{G}}$ can be modeled as a realization from the sample space $F = \left\{ {{\boldsymbol{h}_{r,k}}\left( \xi  \right),\boldsymbol{G}\left( \xi  \right),\forall k,\forall \xi } \right\}$ dominated by the knowledge of the imperfect CSI and the channel estimation errors' distribution, where $\xi $ denotes the index of the random realizations drawn from $F$.
  
  In this setup, we formulate the optimization problem in the imperfect CSI case as 
   \begin{equation}   \label{xu9}
  \left.\begin{array}{l}
  {\rm{P}}2:\mathop {\max }\limits_{{{\boldsymbol{W}\left( \xi  \right),\boldsymbol{\theta} \left( \xi  \right)}} } {f_B}\left( {\boldsymbol{W},\boldsymbol{\theta} } \right) = {\mathbb{E}_\xi }\left[ \sum\limits_{k = 1}^K {{q _k}} \log_2 \left( {1 + {\gamma _k}\left( \xi  \right)}  \right)\right]\\
  \;\;\;\;\;\;\;\;\;\;\;\;s.t.\;\;\;\;\left| {{{\theta} _n}\left( \xi  \right)} \right| = 1,\;\;\;\;\;\;\;{\forall n} = 1,2, \cdots ,N,\\
  \;\;\;\;\;\;\;\;\;\;\;\;\;\;\;\;\;\;\;\;\;\angle {{\theta} _n}\left( \xi  \right) \in S,\;\;\;\;\;\;\;{\forall n} = 1,2, \cdots ,N,\\
  \;\;\;\;\;\;\;\;\;\;\;\;\;\;\;\;\;\;\;\;\;\sum\limits_{k = 1}^K {{{\left\| {{\boldsymbol{w}_k}\left( \xi  \right)} \right\|}^2}}  \le {P_t}, \forall \xi .
  \end{array}\right.
  \end{equation}
  
The model of P2 is inspired by \cite{guo2020weighted}. The difference is that \cite{guo2020weighted} decomposed the non-convex optimization problem into two parts: active beamforming with perfect CSI and passive beamforming with imperfect CSI. While P2 considers both imperfect CSI at the BS and RIS. P2 is a stochastic optimization problem with the expectation operator, which is complex to solve because of the complicate probability density function of the sample space $F$.

\section{Proposed Deep Quantization Neural Network for Passive Beamforming}
To address the above problems, we propose a deep quantization neural network framework, called the DQNN.  The DQNN has a two-stage design, offline training and online prediction, which are not the same with the normal DL methods. During the offline training, we train the DQNN with training samples consisting of estimated channels. After the training is completed, we replace the scalar quantization layer with a real quantization layer whose outputs are real phases in the set $S$ during the online prediction. Then, we explore the parameter settings and propose the I-DQNN to adapt to different resolutions of discrete phase shifters. Finally, we also extend DQNN and I-DQNN to the P2. Both DQNN and I-DQNN are unsupervised learning without labels of the optimal RIS reflection vector and transmit precoding matrix.

 \subsection{The Structure of DQNN}
\begin{figure*}[tbp]
 	\centering {
 		\begin{tabular}{ccc}
 			\includegraphics[width=1\textwidth]{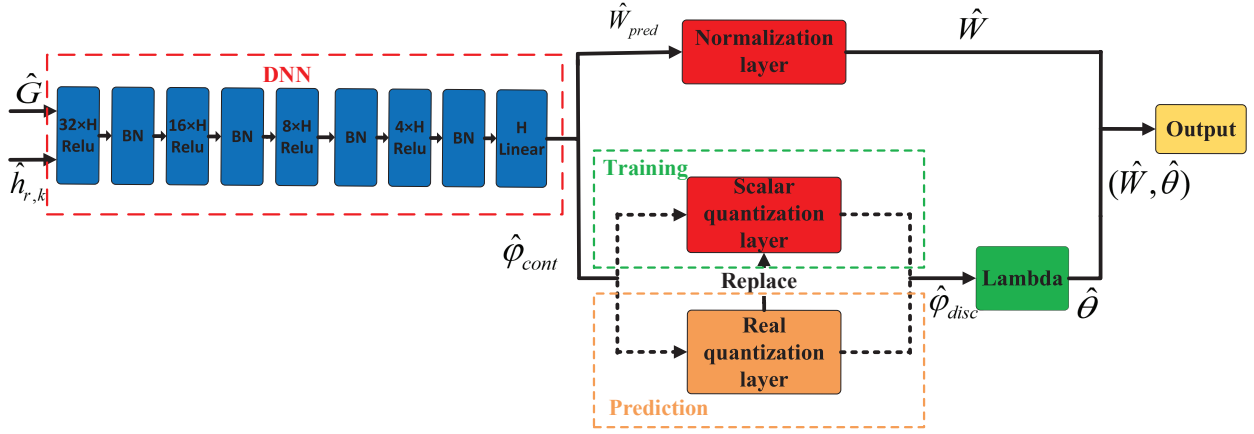}
 		\end{tabular}
 	}
 	\caption{The structure of DQNN.}
 	\vspace{-0.45\baselineskip}
 	\label{fig3}
 \end{figure*}

 The structure of DQNN is illustrated in Fig. \ref{fig3}, consisting of a DNN, a normalization layer, a scalar quantization layer, a real quantization layer, and a lambda layer. The inputs of DQNN are the estimated channels $\hat {\boldsymbol{h}}_{r,k}$ and $\hat {\boldsymbol{G}}$. Because (\ref{xu10}) shows that the active transmit beamforming $\boldsymbol{W}$ and the reflection coefficient vector $\boldsymbol{\theta}$ can be obtained by a function of $\hat {\boldsymbol{h}}_{r,k}$ and $\hat {\boldsymbol{G}}$.
 \begin{equation}   \label{xu10}
 \left.\left( {\boldsymbol{\theta} ,\boldsymbol{W}} \right)  = f( {{\hat {\boldsymbol{h}}_{r,k}},{\hat {\boldsymbol{G}}}} ).\right.
 \end{equation}

 The universal approximation theorem shows that a feedforward neural network with only a single hidden layer and a finite number of neural units can fit functions of arbitrary complexity with arbitrary precision \cite{1991Approximation}. Therefore, the DNN can learn the mapping function shown in (\ref{xu10}). Note that the elements of $\boldsymbol{W}$ and $\boldsymbol{\theta}$ are all complex. However, the neural network is based on real values. Thus, we do not directly learn the mapping of the estimated channels $\hat {\boldsymbol{h}}_{r,k}$ and $\hat {\boldsymbol{G}}$ to the complex $\boldsymbol{W}$ and $\boldsymbol{\theta}$. Instead, we learn the vector consisting of the real and imaginary parts of the $\boldsymbol{W}$ and the real phases of the $\boldsymbol{\theta}$, which is different from that many DL-based methods learn the real and imaginary parts of $\boldsymbol{\theta}$ as well as. Because the real phase ${{\hat {\boldsymbol{\varphi}} }_{cont}}$ is easier to be processed by our proposed scalar quantization layer than the real and imaginary parts of $\boldsymbol{\theta}$. More specifically, we use a DNN consisting of four batch normalization (BN) layers and five fully-connected layers with the numbers of neurons of $32H$, $16H$, $8H$, $4H$, and $H$ to output the real and imaginary parts of ${{\hat {\boldsymbol{W}}}_{pred}}$ and the ${{\hat {\boldsymbol{\varphi}} }_{cont}}$, where the $H = N + 2KM$ is the dimension of the outputs. The last layer uses a linear unit, while other layers use the rectified linear unit (Relu) activation function. 
 
 
 The ${{\hat {\boldsymbol{\varphi}} }_{cont}}$ and ${\hat{\boldsymbol{\varphi}}_{disc}}$ are the input and output, respectively, of the scalar quantization layer and the real quantization layer. For the unit-modulus constraint in P1, we use a lambda layer to convert the real-value phase ${\hat{\boldsymbol{\varphi}}_{disc}}$ into a complex reflection coefficient $\hat{\boldsymbol{\theta}}$, which has a unit modulus. The function of the lambda layer is given by
 
 \begin{equation}   \label{xu11}
 \left.{\hat{\boldsymbol{\theta}} } = {e^{j \cdot {{\hat{\boldsymbol{\varphi}}_{disc}}}}} = \cos \left( {{{\hat{\boldsymbol{\varphi}}_{disc}}}} \right) + j \cdot \sin \left( {{{\hat{\boldsymbol{\varphi}}_{disc}}}} \right).\right.
 \end{equation}
 
 To implement the power constraint defined in (\ref{xu5}), a normalization layer is employed at the outputs ${\operatorname{Re}({\hat {\boldsymbol{W}} }_{pred})}$ and ${\operatorname{Im}({\hat {\boldsymbol{W}} }_{pred})}$ of the DNN. The function of the normalization layer is expressed as
 
  \begin{equation}   \label{xu12}
 \left.\hat {\boldsymbol{W}} = \sqrt {{P_t}} \frac{{{\operatorname{Re}}\left( {{{\hat {\boldsymbol{W}}}_{pred}}} \right) + j \cdot {\operatorname{Im}}\left( {{{\hat {\boldsymbol{W}}}_{pred}}} \right)}}{\sqrt {{{{\left\| {{\operatorname{Re}}\left( {{{\hat {\boldsymbol{W}}}_{pred}}} \right)} \right\|}^2} + {{\left\| {{\operatorname{Im}}\left( {{{\hat {\boldsymbol{W}}}_{pred}}} \right)} \right\|}^2}}}}.\right.
 \end{equation}

 Then we use ${\hat{\boldsymbol{\theta}}} $, $\hat {\boldsymbol{W}}$ to calculate the loss function for P1, which can be expressed as
 \begin{equation}   \label{xu13}
 \left.Los{s_{1}} =  - \sum\limits_{l = 1}^L {\sum\limits_{k = 1}^K {{q_{k,l}}} {{\log }_2}\left( {1 + {\gamma _{k,l}}} \right)}  \right.
 \end{equation}
 where $L$ is the batch size of the training samples. The loss function represented in (\ref{xu13}) is the negative sum of the WSR of a batch, which means that the maximum WSR can be obtained by minimizing the loss function.
 
 \subsection{Scalar Quantization Layer}
 In this subsection, we introduce the scalar quantization layer inspired by \cite{22xu}. While considering using DL-based methods to solve P1, one of the most critical challenges is the constraint of discrete phase shifts. The key idea of DQNN to address this challenge is that we first obtain the continuous phases ${{\hat {\boldsymbol{\varphi}} }_{cont}}$ by relaxing the discrete phase shifts constraint to the continuous phase shifts constraint. Secondly, we employ a continuous-to-discrete mapping called quantization layer, which can quantize the continuous phases to the discrete phases in set $S$. 
 
 The essence of establishing a quantization layer in DQNN is to use a quantization function as the continuous phases' activation function. The quantization function can be modeled as a superposition of some step functions to map the continuous regions into a single value. However, this is difficult to achieve in stochastic gradient descent (SGD)-based DL when optimizing the DQNN. Because the step function is non-differentiable, the gradient cannot be passed back to the pre-network through the quantization function. 

 To address this challenge, we construct a scalar quantization layer based on a soft-to-hard quantization technique, which is based on approximating the non-differentiable quantization function by a differentiable one \cite{2017arXiv170400648A}. To be precise, we use a sum of shifted hyperbolic tangents as the activation function, which is differentiable, to replace the non-differentiable quantization function. Note that other differentiable activation functions can also be used, such as sigmoid and arctangent. They have similar property as the shifted hyperbolic tangent by controlling their steepness, amplitude, quantization area, and other characteristics. 
 
 It can be seen from Fig. \ref{fig2} that shifted hyperbolic tangents can approximate the quantization function very well, and it is a differentiable function that is easy to be applied in SGD-based DL. To limit the output phases to $[0,2\pi )$, we also make a little adjustment to the non-linear function, which can be expressed as
 \begin{align} \label{xu14}
 {Q_A}\left( x \right) = \sum\limits_{i = 1}^{B - 1} {{a_i}\left[ {\tanh \left( {{c_i}\left( {x - {{\rho}_i}} \right)} \right) + 1} \right]},
 \end{align}
 where $\left\{ {{a_i},{{\rho}_i},{c_i}} \right\}, i = 1, 2,..., B - 1$, is the set of real-valued parameters of ${Q_A}\left( x \right)$. $\left\{ {{a_i}} \right\}$ are the amplitudes, which represent the phases of $\boldsymbol{\theta}$ after quantization. Note that the phases of $\boldsymbol{\theta}$ are quantized uniformly, depending on $b$, the number of the quantization bits. Therefore, $\left\{ {{a_i}} \right\}$ are also determined by $b$ and set as ${a_i} =a= \frac{\pi }{{{2^b}}}$.  $\left\{ {{{\rho}_i}} \right\}$ represent the quantization areas of the quantization function, and $\left\{ {{c_i}} \right\}$ represent the degree to which the sum of shifted hyperbolic tangents approaches the real non-differentiable quantization function. The larger $\left\{ {{c_i}} \right\}$ are, the closer to the real quantization function. Furthermore, we assume that all $\left\{ {{c_i}} \right\}$ are fixed and can be represented as $c$. The $c$ is regarded as a parameter that can be controlled manually rather than a learnable parameter. During the offline training, the only $\left\{ {{{\rho}_i}} \right\}$ can be used as the learnable parameters.

 As shown in Fig. \ref{fig2}, the abscissa represents the continuous phases output by the DNN, and the ordinate represents the final quantization values determined by $a$. Meanwhile, $\left\{ {{{\rho}_i}} \right\}$ determine the quantization areas corresponding to different quantization values on the abscissa. As the training progresses, $\left\{ {{{\rho}_i}} \right\}$ and the network's parameters will be learned to achieve the final optimization goal.

\subsection{Offline Training and Online Prediction}
DQNN consists of two different stages, offline training and online prediction. One novelty in this work is to make the network structures and strategies different in offline training and online prediction. In the offline training, DQNN uses the scalar quantization layer as a continuous-to-discrete approximate mapping. However, the phases obtained in this way are still continuous. Therefore, we also need to quantize the scalar quantization layer's continuous outputs to discrete phases during the online prediction according to the practical phase shifter's resolution.
 \begin{figure}[tbp]
	\centering {
		\begin{tabular}{ccc}
			\includegraphics[width=0.6\textwidth]{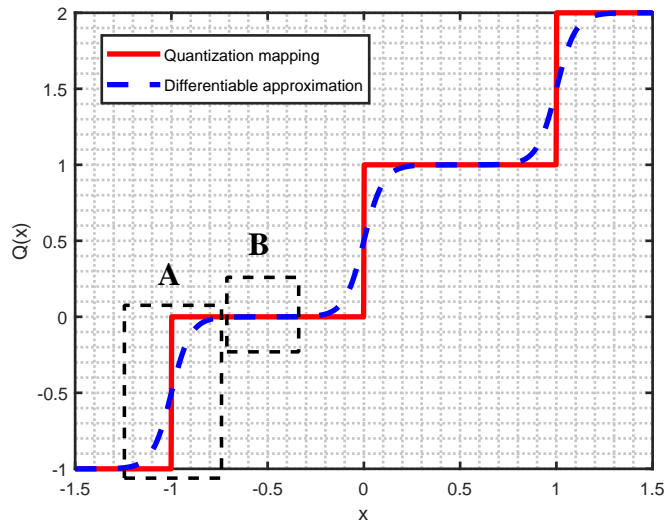}
		\end{tabular}
	}
	\caption{Soft-to-hard quantization.}
	\vspace{-0.45\baselineskip}
	\label{fig2}
\end{figure}

To achieve the above requirements, we use a real quantization layer to replace the approximate scalar quantization layer in the online prediction, which can be expressed as
 \begin{equation}   \label{xu15}
\left.{Q_R}\left( x \right) = \left\{ \begin{array}{l}
0\;\;\;\;\;\;\;\;\;\;\;\;\;\;\;\;\;\;\;\;\;\;\;\;\;\;\;\;x < {{\rho}_0}\\
{Q_A}\left( {\frac{{{{\rho}_i} + {{\rho}_{i + 1}}}}{2}} \right)\;\;\;\;\;\;\;\;{{\rho}_i} \le x < {{\rho}_{i + 1}}\\
\sum\limits_{i = 1}^{B - 1} {2 \cdot {a_i}\;\;\;\;\;\;\;\;\;\;\;\;\;\;\;\;{{\rho}_{B - 1}} \le x\;}
\end{array} \right\},\right.
\end{equation}
where ${{\rho}_0} \le {{\rho}_1} \le  \cdots  \le {{\rho}_{B - 1}}$ (when this condition is not satisfied, we re-sort and re-index the parameters according to the size order), and ${{{\rho}_i}}$ determines the decision region of each step function in the scalar quantization layer, which is because when $\tanh \left( {{c_i}\left( {x - {{\rho}_i}} \right)} \right) = 0$, we can get $x = {{\rho}_i}$ as the decision region. 

\begin{table}[tbp]\normalsize
	\caption{\label{A1}} 
	\begin{tabular}{lcl} 
		\toprule 
		Algorithm 1: The DQNN-based algorithm  \\ 
		\midrule 
		\textbf{Input:} $\hat {\boldsymbol{h}}_{r,k}$, $\hat {\boldsymbol{G}}$, ${q _k}$, ${\sigma ^2}$.\\
		\textbf{Output:} $\hat {\boldsymbol{\theta}}$, $\hat {\boldsymbol{W}}$.\\
		1. Choose a random $c$, set $i=1$ and ${\rm{WSR}}_p^{\left( 0 \right)}=0$; \\
		2. Train the DQNN with training samples and predict 
		the ${\rm{WSR}}_t^{\left( i \right)}$ of validation samples;\\  
		3. Replace the scalar quantization layer with real quantization layer (\ref{xu15}) to predict the \\${\rm{WSR}}_p^{\left( i \right)}$ of validation 
		samples and save model;\\
		4. Calculate the gap ${G_i}$ between ${{\rm{WSR}}_t^{\left( i \right)}}$ and ${{\rm{WSR}}_p^{\left( i \right)}}$ 
		by (\ref{xu16});\\
		5. If ${\rm{WSR}}_p^{\left( i \right)} \ge {\rm{WSR}}_p^{\left( i-1 \right)}$:\\
		6.\qquad If ${G_i} < \tau$:\\
		7.\qquad\qquad $c=c-1,i=i+1$;\\
		8.\qquad Else:\\ 
		7.\qquad\qquad $c=c+1,i=i+1$;\\
		8.\qquad Skip to step 2;\\
		8. Else:\\ 
		9.\qquad Return model;  \\
		10. Do step 2 and 3 with test samples to output $\hat {\boldsymbol{\theta}}$, $\hat {\boldsymbol{W}}$;\\  
		\bottomrule 
	\end{tabular} 
\end{table} 

The function $Q_R$ quantizes the continuous phases to the corresponding discrete phases according to the real quantization resolution and the scalar quantization layer parameters learned in the offline training. Note that the continuous function $Q_A$ is an approximation of the quantization function $Q_R$. When the $Q_R$ is used to replace the $Q_A$, there is a gap between their outputs ${\hat{\boldsymbol{\varphi}}_{disc}}$, resulting in a decrease in the corresponding WSR. This gap is controlled by $c$. It can be seen from Fig. \ref{fig2} that the steep region A of the quantization function will become smaller as $c$ increases, which indicates that function ${Q_A}$ is getting closer to function ${Q_R}$. Moreover, it will reduce the gap of WSR, which means that the performance of online prediction will be consistent with the performance of offline training. However, the smaller the region A is, the greater the restriction on the inputs of the scalar quantization layer is. That will increase the difficulty of network training and convergence. On the contrary, when the learning ability of the neural network is sufficient and $c$ is small, a large amount of ${\hat{\boldsymbol{\varphi}}_{cont}}$ may be distributed in region A, which will cause the gap to become larger. Even if we get the optimal WSR in the training stage, it is also possible to obtain a small WSR in the online prediction because of this gap. Therefore, one of the keys for DQNN to obtain better performance is to choose the optimal parameter $c$. 

To obtain the optimal $c$, we propose a comparative search method. More specifically, we initialize  $c$ randomly at first. Secondly, we use the initial $c$ to train the DQNN with training samples. After the training is completed, the ${\rm{WSR}}_t$ of the validation samples is obtained. Thirdly, we save this model and predict the reflecting elements' continuous phases of validation samples. Then we use $Q_R$ to quantize the continuous phases to discrete and calculate the corresponding ${\rm{WSR}}_p$. We also need to calculate the gap between ${\rm{WSR}}_p$ and ${\rm{WSR}}_t$, which is defined as 
\begin{equation}   \label{xu16}
\begin{aligned}
{G_i} = \frac{{{\rm{WSR}}_t^{\left( i \right)} - {\rm{WSR}}_p^{\left( i \right)}}}{{{\rm{WSR}}_t^{\left( i \right)}}}.
\end{aligned}
\end{equation} 

When the gap ${G_i}$ is less than our preset threshold $\tau$, the ${{\hat {\boldsymbol{\varphi}} }_{cont}}$ are almost distributed in flat region B, shown in Fig. \ref{fig2}. Then, we need to reduce $c$ for a possible more sufficient training. This would make better performance with a tradeoff between the gap and the training effect. Otherwise, increase $c$ for a smaller gap. Then repeat the above process until the ${\rm{WSR}}_p$ no longer increases. The details of the DQNN-based algorithm are shown in TABLE \ref{A1}.

\subsection{I-DQNN-based Passive Beamforming}
\begin{figure*}[tbp]
	\centering {
		\begin{tabular}{ccc}
			\includegraphics[width=1\textwidth]{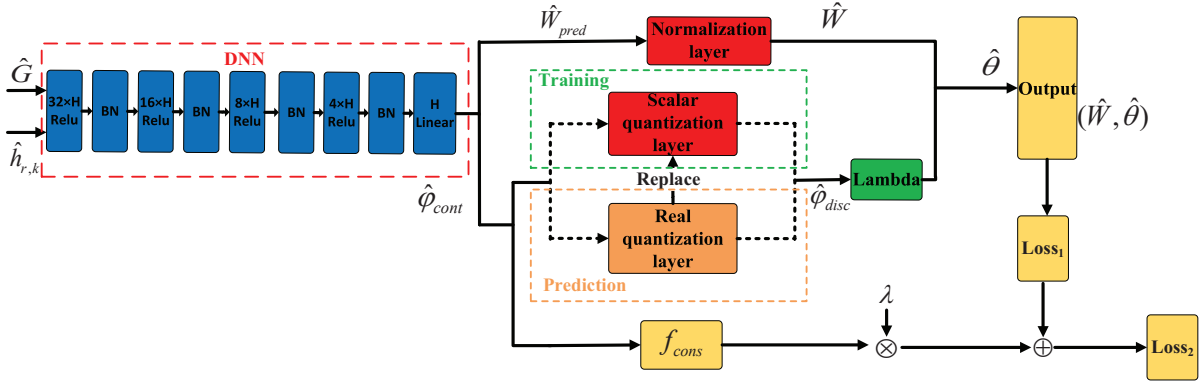}\\
		\end{tabular}
	}
	\caption{The structure of I-DQNN.}
	\vspace{-0.45\baselineskip}
	\label{fig4}
\end{figure*}

In this subsection, we propose an improved network structure, I-DQNN, shown in Fig. \ref{fig4}, based on DQNN.

As mentioned above, DQNN relies on $c$ to control the gap between the WSR of offline training and online prediction. However, the process to determine $c$ by the comparative search method is complicate somehow.

Compared with DQNN, we add a gradient-based constraint derived from (\ref{xu14}) to I-DQNN to solve the problems as mentioned above. The constraint is defined as
 \begin{equation}   \label{xu17}
\left.{f_{cons}} = \sum\limits_{i = 1}^{B - 1} {\frac{{4 \cdot {a_i} \cdot {c_i}}}{{{{\left\{ {{e^{\tanh \left[ {{c_i}\left( {x - {{\rho}_i}} \right)} \right]}} + {e^{\tanh \left[ { - {c_i}\left( {x - {{\rho}_i}} \right)} \right]}}} \right\}}^2}}}}.\right.
\end{equation}

\begin{figure}[tbp]
	\centering {
		\begin{tabular}{ccc}
			\includegraphics[width=0.6\textwidth]{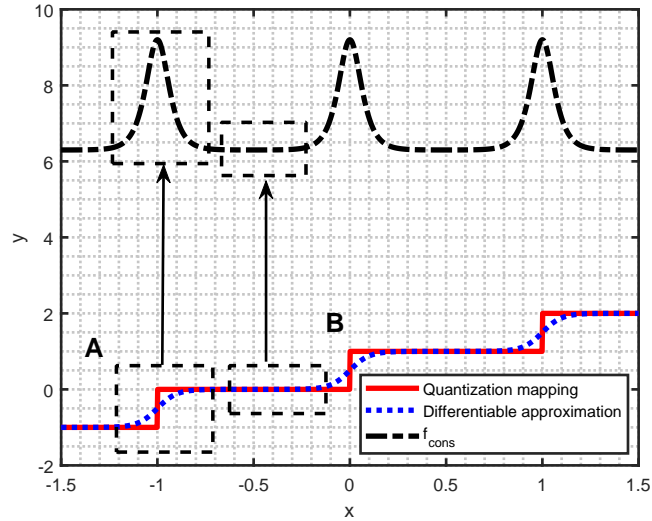}\\
		\end{tabular}
	}
	\caption{The gradient-based constraint ${f_{cons}}$ of I-DQNN.}
	\vspace{-0.45\baselineskip}
	\label{result5}
\end{figure}

${f_{cons}}$ is based on the gradient of ${Q_A}$ for $x$ which is represented in Fig. \ref{result5}. We use two tanh functions to limit the value of ${{c_i}\left( {x - {{\rho}_i}} \right)}$ and ${-{c_i}\left( {x - {{\rho}_i}} \right)}$, respectively, and also to avoid the exponent being too large, which is harmful to train. The detailed derivations are shown in Appendix A.

We construct the loss function of I-DQNN as 
 \begin{equation}   \label{xu18}
\left.Los{s_2} = Los{s_1} + \sum\limits_{l = 1}^L {\lambda  \cdot {f_{cons,l}}} \right.,
\end{equation}
which consists of the negative of weighted sum rate and the penalty term $\lambda  \cdot {f_{cons}}$. During training, our goal is to minimize the loss function, which is equivalent to maximizing the weighted sum rate and minimizing the penalty term. Maximizing the weighted sum rate is our original optimization objective, and minimizing the penalty term aims to eliminate the gap between offline training and online prediction. In I-DQNN, the ${\hat{\boldsymbol{\varphi}}_{cont}}$ are used as the input $x$ of ${f_{cons}}$. The smaller ${f_{cons}}$  means that more ${\hat{\boldsymbol{\varphi}}_{cont}}$ fall into the quantization function's flat region B. There will be no big gap problem when online prediction uses a real quantization layer to replace the scalar quantization layer. 
	
	Conversely, when ${f_{cons}}$ is bigger, more ${\hat{\boldsymbol{\varphi}}_{cont}}$ are in the steep region A of the quantization function, resulting in a more severe gap. Therefore, we add ${f_{cons}}$ to the loss function so that ${\hat{\boldsymbol{\varphi}}_{cont}}$ can be limited and distributed in region B as much as possible. $\lambda$ is the coefficient of the penalty term, and its role is to strike the balance of the significance of these two parts in the loss function. In the following, we propose a heuristic method to determine the value of $\lambda$. Firstly, we set $c=1$ as the initial value. Secondly, this initialized $c$ is applied to the pre-training of I-DQNN to get the corresponding ${{WSR_c}}$ and ${{f_{cons,c}}}$. Noting that, in this pre-training process, the loss function is not (\ref{xu18}) instead (\ref{xu13}). Because, our purpose is to find the ${{f_{cons,c}}}$ with the optimal ${{WSR_c}}$ for the determination of $\lambda$, which can be calculated by 
 \begin{equation}   \label{xu19}
\left.\lambda  = \frac{{0.1 \times {WSR_c}}}{{{f_{cons,c}}}}.\right.
\end{equation}

In the DQNN parameter optimization process, we assume that $K$ times comparative searches are required to get the best parameter $c$. Therefore, we need to repeat $K$ times step 2 and step 3, including $K$ times offline training and $K$ times online predictions of validation samples with the real quantization layer. In the I-DQNN parameter optimization process, only one-time pre-training of I-DQNN is necessary to obtain ${{WSR_c}}$ and ${f_{cons}}$ to calculate the optimal parameter $\lambda$, which means that one-time offline training and online prediction are required. Therefore, we can find that the parameter optimization process of I-DQNN is much simpler and determined than DQNN. 

The training process and prediction process of I-DQNN are the same as DQNN. The real quantization layer also needs to be used in the online prediction  to replace the scalar quantization layer in the offline training. The details of I-DQNN-based algorithm are shown in TABLE \ref{A2}.

\begin{table}[tbp]\normalsize
	\caption{\label{A2}} 
	\begin{tabular}{lcl} 
		\toprule 
		Algorithm 2: The I-DQNN-based algorithm  \\ 
		\midrule 
		\textbf{Input:} $\hat {\boldsymbol{h}}_{r,k}$, $\hat {\boldsymbol{G}}$, ${q _k}$, ${\sigma ^2}$.\\
		\textbf{Output:} $\hat {\boldsymbol{\theta}}$, $\hat {\boldsymbol{W}}$.\\
        1. Initial $c$ as 1.\\
		2. Pre-training I-DQNN with the initial $c$ for corresponding
		${{WSR_c}}$ and ${{f_{cons,c}}}$;\\
		3. Use ${{WSR_c}}$ and ${{f_{cons,c}}}$ obtained from step 2 to calculate
		$\lambda$ by (\ref{xu19});\\
		4. Train I-DQNN with the $\lambda$ and other inputs;\\
		5. Replace scalar quantization layer with real quantization 
		layer (\ref{xu15}) to predict the $\hat {\boldsymbol{\theta}}$ and $\hat {\boldsymbol{W}}$ \\of test samples \\
		\bottomrule 
	\end{tabular} 
\end{table}

\subsection{Extend the DL-based Algorithms for The Imperfect CSI Setup }
In this subsection, we extend our proposed DL-based algorithms to solve P2 with the imperfect CSI setup.

It can be seen that P2 is a challenging problem due to the expectation operator. In \cite{guo2020weighted}, the expectation operator can be solved by the sample average approximation method and stochastic SCA technology, which needs to choose different samples from the space $F$ in each iteration. However, this method needs many iterations to ensure enough samples are taken from $F$ in the practical online application. Instead, this sampling process can be easily transplanted to the offline training process of our proposed algorithms. Because the distribution of the channel estimation error can be learned from the large amount of channel training data collected, which does not cause extra complexity in the practical online application. 

More specifically, we generate the input training data set with the estimated CSI at first. Then, we obtain enough samples from the corresponding $F$ to build the average loss functions of the DQNN and I-DQNN in the offline training. Two average loss functions can be expressed in the same form, 
	\begin{equation}   \label{new1}
	\left.AveLos{s_i} = \sum\limits_j^J {Los{s_{i,j}}} ,\;\;\;\;\;i = 1,2,  \right.
	\end{equation}
	where $J$ is the number of the samples obtain from $F$, $i=1$ is for DQNN, $i=2$ is for I-DQNN, ${Los{s_{i,j}}}$ represents the original loss function of the $j$-th sample from $F$. 

The offline training process of the imperfect CSI case is the same as that of the perfect case, except that the loss function is replaced by (\ref{new1}). After the offline training is completed, the online prediction process is the same as the perfect CSI case without a complex sampling process. 
	
\section{Simulation results}
\begin{figure}[tbp]
	\centering {
		\begin{tabular}{ccc}
			\includegraphics[width=0.6\textwidth]{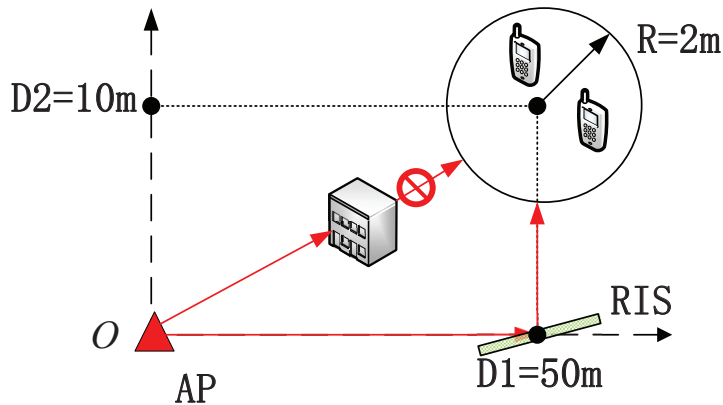}\\
		\end{tabular}
	}
	\caption{The simulated RIS-aided MISO communication scenario comprising of one $M$-antenna AP, $K$ users and one $N$-element RIS.}
	\vspace{-0.45\baselineskip}
	\label{fig5}
\end{figure}

This section presents the simulation results to demonstrate the effectiveness of the proposed DL-based algorithms in the joint beamforming task of the RIS-assisted multiuser MISO system with discrete phase shifts and imperfect CSI. The details of the system refer to \cite{guo2020weighted} and are illustrated in Fig. \ref{fig5}, in which one AP equipped with 4 antennas locates at $\left( {0,0} \right)$, and 2 users with a single antenna randomly distributed in a circle centered at $\left( {50,10} \right)$ with radius 2 m under the Cartesian coordinate system in meter (m). We assume that the RIS with a uniform rectangular array (URA) is located at $\left( {50,0} \right)$. We set threshold $\tau = 0.005$.

We consider the large-scale fading and the small-scale fading to describe the real channel environment. Specifically, the path loss of large-scale fading can be expressed as $\beta \left( d \right) = {\beta _0}{\left( {\frac{d}{{{d_0}}}} \right)^{ - p }}$, where $d$ represents the distance between the individual links, ${\beta _0}=-35.6$ dB is the corresponding channel gain when the reference distance ${d_0} = 1$ m, and $p $ represents the path loss exponent. We set $p=2.2$ for the channel AP-RIS and RIS-user. For small-scale fading, we use the Rician channel model for two individual links, expressed as 
 \begin{equation}   \label{xu20}
\left.{\boldsymbol{h}_{r,k}} = \sqrt {\frac{{{\kappa _{r,k}}}}{{1 + {\kappa _{r,k}}}}} \boldsymbol{h}_{r,k}^{LoS} + \sqrt {\frac{1}{{1 + {\kappa _{r,k}}}}} \boldsymbol{h}_{r,k}^{NLoS},\right.
\end{equation}
 \begin{equation}   \label{xu21}
\left.{\boldsymbol{G}} = \sqrt {\frac{{{\kappa _{G}}}}{{1 + {\kappa _{G}}}}} \boldsymbol{G}^{LoS} + \sqrt {\frac{1}{{1 + {\kappa _{G}}}}} \boldsymbol{G}^{NLoS},\right.
\end{equation}
where ${{\kappa _{G}}}$ and ${{\kappa _{r,k}}}$ are the Rician factor of AP-RIS link and RIS-user link, respectively. We set ${{\kappa _{r,k}} = {\kappa _{G}} = 10}$. $\boldsymbol{h}_{r,k}^{LoS}$ and $\boldsymbol{G}^{LoS}$ are the Line of Sight (LoS) components, $\boldsymbol{h}_{r,k}^{NLoS}$ and $\boldsymbol{G}^{NLoS}$ are the None Line of Sight (NLoS) components, repectively. 

For the different setups of perfect CSI and imperfect CSI, we generate 200000, 1000 and 1000 samples for training, validation and prediction, respectively. In particular, all imperfect CSI samples are the estimated channels with errors compared to the real channels. The error can be expressed as

 \begin{equation}   \label{x22}
\left.\eta  = \frac{{\mathbb{E}{\left[ {{{\left| {x - \hat x} \right|}^2}} \right]}}}{{\mathbb{E}{\left[ {{{\left| {\hat x} \right|}^2}} \right]}}},\right.
\end{equation}
where $x$ is one element of the channels, and $\hat x$ is the corresponding estimate value. We assume that the estimate error $x-{\hat x}$ follows zero mean complex Gaussian distribution, and all these elements have the same normalized MSE $\eta$. Noting that, $\eta=0$ for the case of perfect CSI. 

We stack the real and imaginary parts of each complex sample into a vector which can be defined as $\left[ {{\mathop{\rm Re}\nolimits} \left\{ \boldsymbol{G} \right\},{\rm Im}\left\{ \boldsymbol{G} \right\},{\mathop{\rm Re}\nolimits} \left\{ {\boldsymbol{h}_{r,k}} \right\},{\rm Im}\left\{ {\boldsymbol{h}_{r,k}}  \right\}} \right]$. Then we take a standardization for this vector as the final inputs of the DQNN and I-DQNN. In the offline training, we use the early stopping strategy with patience 50 to reduce overfitting and set the number of maximal epoch to 1500. Moreover, we set the initial learning rate as 0.001 and when the loss of the validation samples does not decrease for 20 consecutive epochs, we reduce the learning rate by a factor of 0.8 until to 0.00005. In all experiments, the batch size is set to 1024.

\subsection{Benchmarks for Comparison}
We compare the performance of the proposed DL-based algorithms with the following benchmark schemes for the cases of perfect CSI and imperfect CSI:

\begin{itemize}
	
	\item \textbf{Scheme 1} (AO algorithm): The algorithm proposed in \cite{8mainYou} for the joint beamforming with discrete phase shift constraint based on the AO and the successive refinement algorithm.
	\item \textbf{Scheme 2} (Extended BCD): For the perfect CSI case, the continuous phase shifts of passive beamforming are firstly obtained by the BCD-based algorithm \cite{guo2020weighted} in each iteration. Then we applied the successive refinement algorithm proposed in \cite{8mainYou} to obtain discrete phase shifts from the optimized continuous phase shifts. For the imperfect CSI case, the continuous phase shifts of passive beamforming are solved by the imperfect CSI version of BCD-based algorithm proposed in \cite{guo2020weighted}, and the other process is the same as the perfect CSI's.
	\item \textbf{Scheme 3} (Random discrete phase shifts): ${\boldsymbol{\theta}}$ is initialized by randomly choosing value in set $S$, and then ${\boldsymbol{W}}$ is optimized by WMMSE for both perfect CSI and imperfect CSI.
	\item \textbf{Scheme 4} (Upper bound): We use the continuous phase shifts version of BCD-based algorithm \cite{guo2020weighted} as the upper bound.
\end{itemize} 

\subsection{Performance Versus $P_t$}
\begin{figure}[tbp]
	\centering {
		\begin{tabular}{ccc}
			\includegraphics[width=0.55\textwidth]{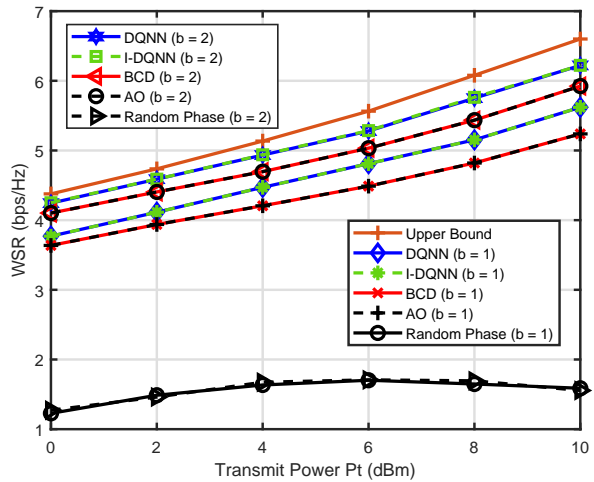}
		\end{tabular}
	}
	\caption{WSR versus transmit power $Pt$ when $N=50$.}
	\vspace{-0.45\baselineskip}
	\label{result1}
\end{figure}

Fig. \ref{result1} shows the WSR of different algorithms versus transmit power $Pt$. In this experiment, we set $N=50$. It is worth mentioning that the results of DQNN are obtained under the optimal $c$ which is fixed and got after a set of comparative search methods proposed in TABLE \ref{T1} with ${N=50}$ and $Pt=5$ dBm for simplicity. For the I-DQNN, we set the initial $c=1$, then the corresponding ${{WSR_c}}$ and ${{f_{cons,c}}}$ can be obtained as 5.225 and 1.382 through simple pre-training of I-DQNN. The $\lambda$ is 0.38 calculated by (\ref{xu19}) and fixed for different ${Pt}$. In Fig. \ref{result1}, we can see that the WSR increases with the transmit power for all algorithms except the random algorithm. Besides, the performance of the extended BCD algorithm, the AO algorithm, DQNN, and I-DQNN all have great improvement compared to the random method, which indicates the importance of a good algorithm for designing RIS. The performance of the extended BCD algorithm and the AO algorithm is almost the same with the same initial points. We also observe that the DQNN and I-DQNN have a very close performance with the same $b$. They have about 1.2 dB and 1 dB gain comparing with the BCD-based algorithm when $b=1$ bit and $b=2$ bits respectively. Moreover, when $b=2$ bits, the performance gap between our proposed DL-based algorithms and the continuous phase shifts case is about 1 dB, which is smaller than 2 dB that is the gap between BCD-based algorithm and the continuous phase shifts case. 

\subsection{Performance Versus $N$}
\begin{figure}[tbp]
	\centering {
		\begin{tabular}{ccc}
			\includegraphics[width=0.55\textwidth]{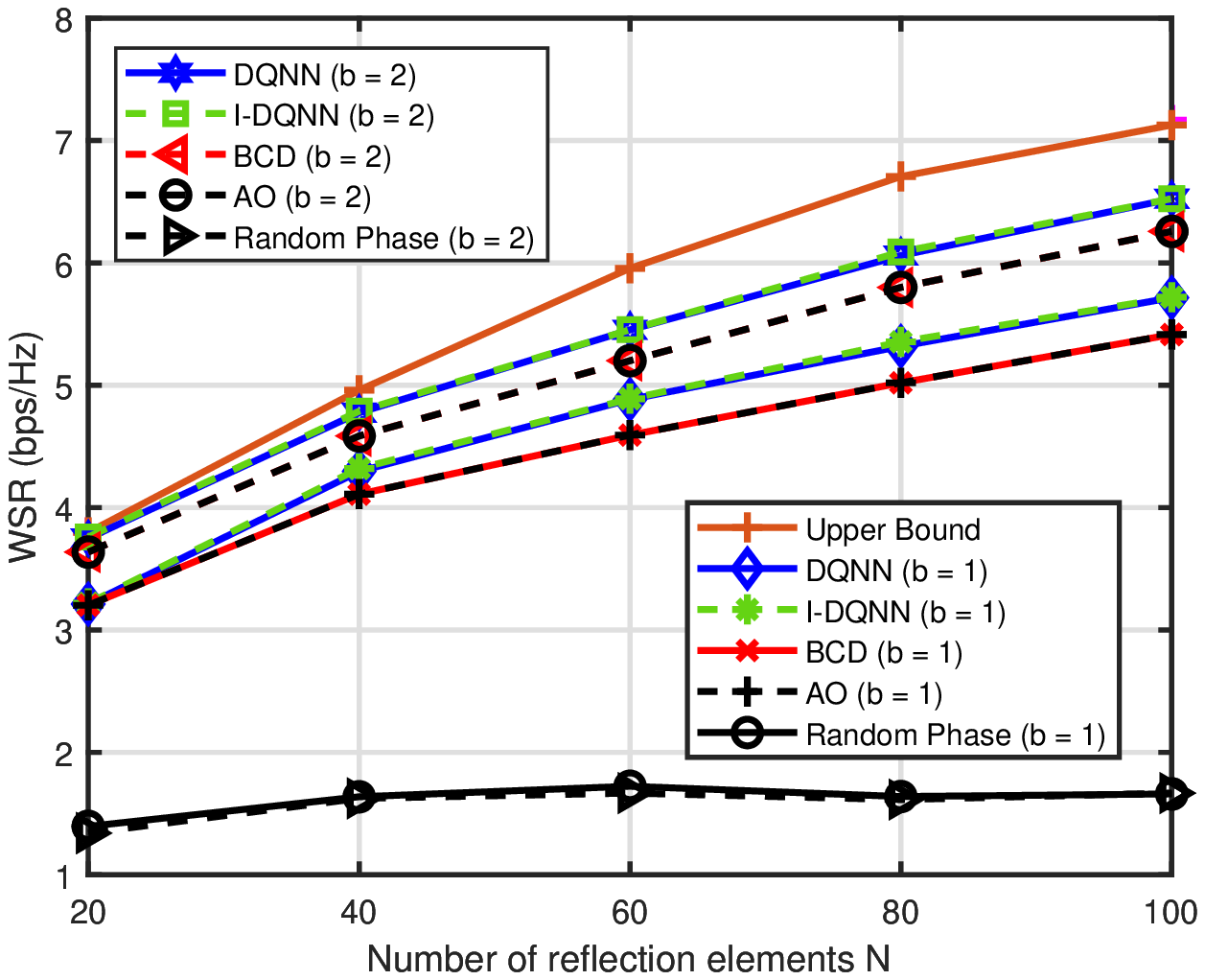}\\
		\end{tabular}
	}
	\caption{WSR versus the RIS size $N$ when $Pt=5$ dBm.}
	\vspace{-0.45\baselineskip}
	\label{result2}
\end{figure}
\begin{figure}[tbp]
	\centering {
		\begin{tabular}{ccc}
			\includegraphics[width=0.55\textwidth]{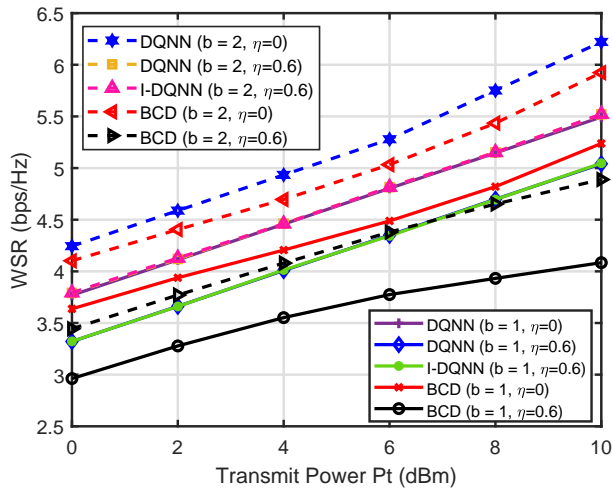}\\
		\end{tabular}
	}
	\caption{WSR versus transmit power $Pt$ when $N=50$, $\eta=0, 0.6$.}
	\vspace{-0.45\baselineskip}
	\label{result3}
\end{figure}
In Fig. \ref{result2}, we fix the transmit power $Pt=5$ dBm and compare the performance of various algorithms versus the size of RIS with the different resolutions of discrete phase shifters. Besides, with different $N$, the optimal $c$ of DQNN obtained by comparative search method and the $\lambda$ calculated by (\ref{xu19}) are illustrated at TABLE \ref{T1} and TABLE \ref{T2} respectively. As we can see that Fig. \ref{result2} shows that the performance of DQNN and I-DQNN are almost the same with different $N$. Compared with the random algorithm, the performance of two DL-based algorithms and BCD-based algorithms is much better since the most reflected signals cannot arrive at the mobile users' receivers with the poor designing of RIS in the random algorithm. Also, we observe that the performance of all algorithms increases with the increase of $N$ except the random algorithm. Moreover, the gap of WSR between DL-based algorithms and BCD-based algorithm is average 0.28 bps/Hz when $b=1$ bit and 0.24 bps/Hz when $b=2$ bits, respectively.

\subsection{Performance Versus $\eta$}
In this subsection, we discuss on the imperfect CSI case which is more complicated to solve in traditional algorithm but simpler in our proposed DL-based algorithms. In Fig. \ref{result3}, we compare the performance of different algorithms to solve P2 versus $Pt$ with the fixed $N=50$. We observe that the performance of all algorithms increases with the increase of $Pt$ under different $\eta$. The difference is that in the BCD-based algorithm, as $Pt$ gradually increases, the performance loss between perfect CSI ($\eta =0$) and imperfect CSI ($\eta =0.6$) gradually increases. For example, when $b=1$ bit and $Pt=6$ dBm, the loss is 5 dB, and when $Pt$ increases to 10 dBm, the loss increases to 8 dB. On the contrary, the performance loss of our proposed DL-based algorithms between perfect CSI ($\eta =0$) and imperfect CSI ($\eta =0.6$) remains almost unchanged with the increase of $Pt$, and it is about 3 dB when $b=1$ bit and $b=2$ bits.    

\begin{figure}[tbp]
	\centering {
		\begin{tabular}{ccc}
			\includegraphics[width=0.55\textwidth]{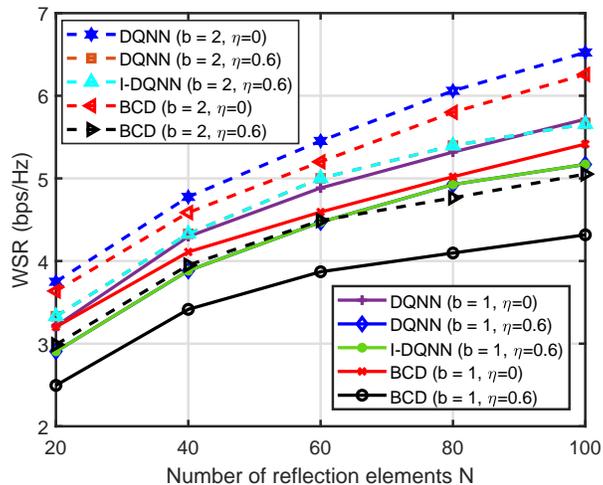}\\
		\end{tabular}
	}
	\caption{WSR versus the RIS size $N$ when $Pt=5$ dBm, $\eta=0, 0.6$.}
	\vspace{-0.45\baselineskip}
	\label{result4}
\end{figure} 
\begin{figure}[tbp]
	\centering {
		\begin{tabular}{ccc}
			\includegraphics[width=0.55\textwidth]{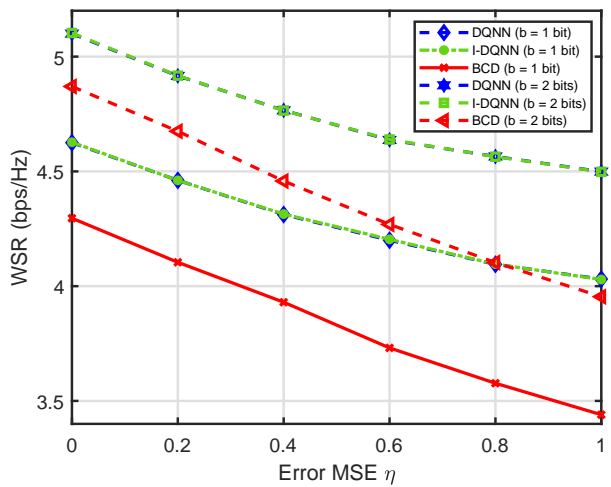}\\
		\end{tabular}
	}
	\caption{WSR versus $\eta$ when $N=50$, $Pt=5$ dBm.}
	\vspace{-0.45\baselineskip}
	\label{result6}
\end{figure}

Next, in Fig. \ref{result4}, we fix the transmit power $Pt=5$ dBm. It can be seen that the performance of all algorithms increases with the increase of $N$ under different $\eta$. For the BCD-based algorithms, the performance loss between perfect CSI ($\eta =0$) and imperfect CSI ($\eta =0.6$) is also increases when $b=1$ bit and $b=2$ bits. For example, when $N=60$ and $b=1$ bit, the gap of WSR between ($\eta =0$) and ($\eta =0.6$) is 0.7 bps/Hz, and it increases to 1 bps/Hz when $N=100$. Similarity, when $b=2$ bits, the gap increases from 0.8 bps/Hz to 1.3 bps/Hz with the same increase of $N$. For the DL-based algorithms, the gap also increases with the increase of $N$ when $b=2$ bits (e.g.,  $WSR=0.5 \sim 1$ bps/Hz, $N=60 \sim 100$). However, when $b=1$ bit, the gap does not change much, with an average of 0.5 bps/Hz. Moreover, it is obviously that our proposed DL-based algorithms have better and more stable performance than BCD-based algorithm.

Finally, in Fig. \ref{result6}, we fix $N=50$ and $Pt=5$ dBm to study the performance of various algorithms with different error MSE $\eta$. It is seen that the performance of all algorithms decreases with the increase of $\eta$. We compare the DL-based algorithms and the BCD-based algorithm with the same $b$. We can see that the performance of the former is much better than the latter and the gap between them increases as the $\eta$ increases. 

Through the above experimental results, we can conclude whether the perfect CSI or imperfect CSI case, our proposed DL-based algorithms' performance is more robust, stable, and better than BCD-based algorithms.

\begin{figure}[tbp]
	\centering
	\subfigure[]{
		\label{FirstStep}
		\begin{minipage}[t]{0.6\linewidth}
			\includegraphics[width=1\textwidth]{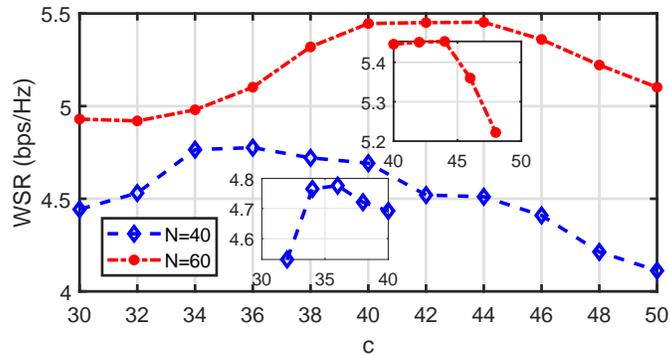}
		\end{minipage}
	}
	\subfigure[]{
		\label{SecondStep}
		\begin{minipage}[t]{0.6\linewidth}
			\includegraphics[width=1\textwidth]{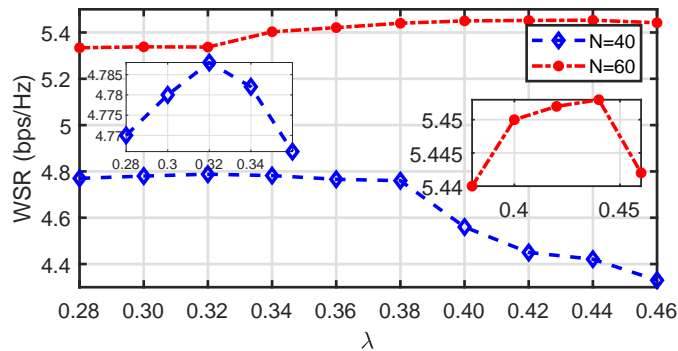}
		\end{minipage}
	}
	\caption{Simulation results of optimal parameters of DQNN and I-DQNN. (a) Achievable rate versus $c$ in DQNN; (b) Achievable rate versus $\lambda$ in I-DQNN.}
	\vspace{-0.5\baselineskip}
	\label{result7}
\end{figure}

\subsection{Comparison Of The Parameters Optimization}

In the above experimental results, we can find that DQNN and I-DQNN have very similar and excellent performance with the premise that they all have the optimal parameters. However, the complexity of determining their optimal training parameters is different. As mentioned above, to get the best results in DQNN, we need to find the optimal $c$, which maximizes the WSR by the time-consuming comparative search method. In contrast, for I-DQNN, the parameter optimization process is much simpler. TABLE \ref{T1} illustrates the optimal $c$ of DQNN. We can observe that $c$ can be easily set as 1 for different $N$ when $b=1$ bit, since it is easy to train with the lower resolution of discrete phase shifters. However, when $b=2$ bits, the training of DQNN is more complicated. Thus the optimal $c$ needs to be determined by the comparative search method. TABLE \ref{T2} shows the optimal parameters with simpler computational process of I-DQNN. Fig. \ref{FirstStep} shows the WSR for different $c$ when $N =40$ and $60$ of DQNN. More specifically, when $N =40$, the $c=36$ obtained by comparative search makes the WSR biggest, and when $N =60$, the optimal $c=44$. This result shows the effectiveness of our proposed comparative search method. Fig. \ref{SecondStep} shows the WSR of I-DQNN with different $\lambda $ and a fixed $c$. When $N =40$, the optimal WSR is 4.776 with the $\lambda = 0.32$, calculated by (\ref{xu19}). Simultaneously, when $N=60$, the calculated $\lambda = 0.44$ can result the optimal WSR of 5.453. This result indicates that the parameter optimization method proposed for the I-DQNN is efficient. Therefore, we can choose DQNN with $c=1$ that has simpler structure than I-DQNN when $b=1$ bit, but choose I-DQNN with a simpler parameter decided process when $b=2$ bits.
\begin{figure}[tbp]
	\centering
	\hspace{-1\baselineskip}
	\subfigure[]{
		\includegraphics[width=0.5\textwidth]{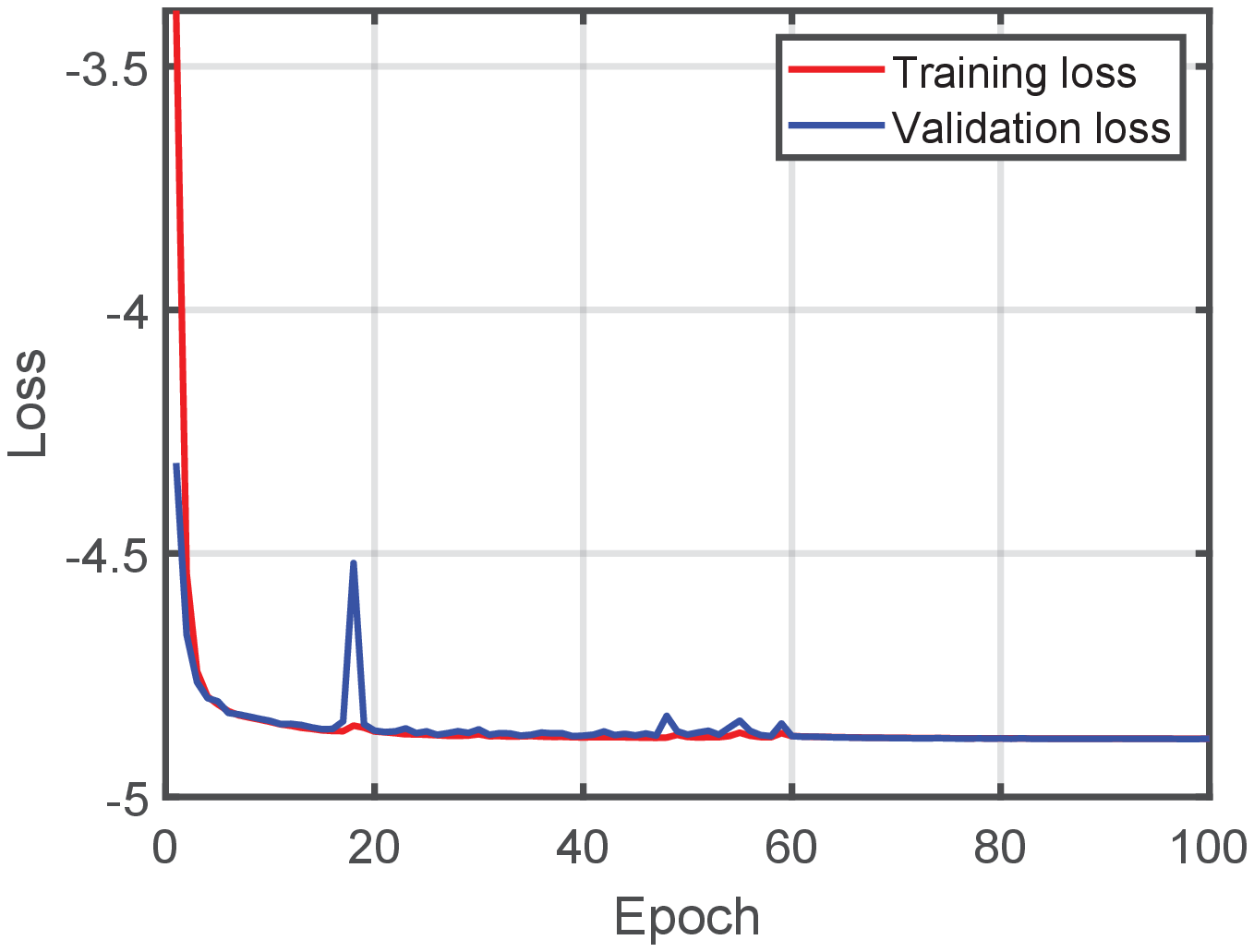}
		\label{Con1}
	}
	\hspace{-1\baselineskip}
	\subfigure[]{
		\includegraphics[width=0.5\textwidth]{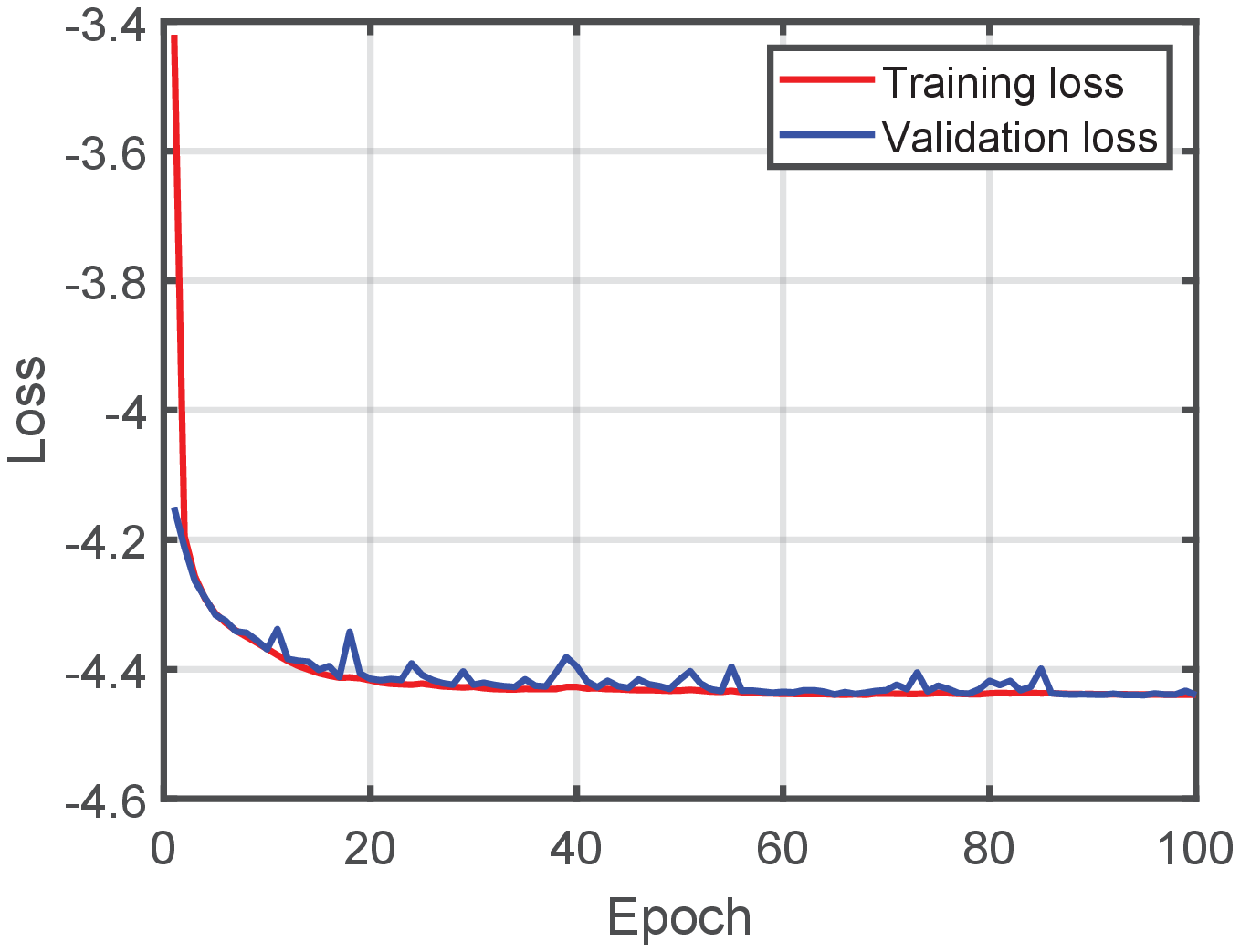}
		\label{Con2}
	}
	\caption{(a) The convergence trend of DQNN. (b) The convergence trend of I-DQNN.}
\end{figure}
\begin{table}[tbp]
	\caption{The optimal $c$ of DQNN versus $N$}
	\vspace{1pt}
	\centering
	\begin{tabular}{cccccc}
		\hline
		\thead[cl]{$N$} & \thead[c]{20}& \thead[c]{40} & \thead[c]{60} & \thead[c]{80} & \thead[c]{100}\\   
		\hline   
		\thead[c]{$c, (b=1)$} & \thead[c]{1}& \thead[c]{1} & \thead[c]{1} & \thead[c]{1} & \thead[c]{1} \\   
		\hline 
		\thead[c]{$c, (b=2)$} & \thead[c]{54}& \thead[c]{44} & \thead[c]{36} & \thead[c]{31} & \thead[c]{29} \\   
		\hline 
	\end{tabular}
	\label{T1}
\end{table}

\subsection{Computational Complexity and Convergence}

In addition to measure the performance of algorithms for solving the non-convex problem, these algorithms' computational complexity also needs to be considered. In \cite{wu2019beamforming}, the authors apply the SDR-based algorithm to the passive beamforming with the fixed active transmit beamforming of MRT. Although it achieves good performance, the complexity of the SDR technique is $O\left( {{N^6}} \right)$, which is very high with a large number of the elements at the RIS. In \cite{guo2020weighted}, the total complexity of the proposed BCD-based algorithm is $O\left( {{I_O}\left( {2KNM + K{M^2} + {K^2}{N^2}} \right)} \right)$, where ${I_O}$ is the number of iterations. Paper \cite{di2020hybrid} shows that the proposed iterative algorithm  for the joint beamforming with discrete phase shifts has a complexity $O\left( {{2^b}{N^2}} \right)$. Although we need much time for the offline training of DQNN, the complexity of online prediction is small. For simplicity, we ignore the calculation of the BN layers and consider the flops of the FC layers as the computational complexity, which is $32H*16H+16H*8H+8H*4H+4H*H=676H^2$, where $H=N+2KM$. Therefore, the total complexity is $O\left( {676({N^2} + 4KMN + 4{K^2}{M^2})} \right)$. It can be concluded that the DQNN-based algorithms have a comparable complexity to the existing state-of-the-art algorithms.
\begin{table}[tbp]
	\caption{The parameters of I-DQNN versus $N$}
	\vspace{1pt}
	\centering
	\begin{tabular}{ccccc}
		\hline
		\thead[c]{$N$} & \thead[c]{initial $c$}& \thead[c]{${{f_{cons,c}}}$} & \thead[c]{${{R_c}}$} & \thead[c]{$\lambda$} \\
		\hline
		20   & 1    & 1.591           & 3.853            & 0.24          \\
		40   & 1    & 1.528           & 4.890            & 0.32           \\
		60   & 1    & 1.294           & 5.702            & 0.44          \\
		80   & 1    & 1.286           & 6.236            & 0.48          \\
		100  & 1    & 1.273           & 6.703            & 0.53           \\  
		\hline       
	\end{tabular}
	\label{T2}
\end{table}

Fig. \ref{Con1} and Fig. \ref{Con2} demonstrate the convergence trends of DQNN and I-DQNN, respectively. We can see that DQNN and I-DQNN converge within 100 epochs, and the training time is short. At the same time, when using the early stopping strategy, the training time would be further shortened.

\section{Conclusion}
In this paper, we investigated the RIS-aided multiuser MISO downlink communication system with discrete phase shifts and imperfect CSI. Firstly, we proposed a DL-based algorithm, DQNN, to solve the joint active transmit beamforming and passive beamforming at the RIS with discrete phase shifts in the perfect CSI case. For the constraint of discrete phase shifts, the DQNN applies a scalar quantization layer to approximate the real quantization function which cannot be achieved in a SGD-based network in offline training, and a real quantization layer to map the continuous phases into the discrete phases during the online prediction. Secondly, we proposed the improved DQNN, I-DQNN, which has a simpler parameter determining process for the more complicated training with the higher bits of discrete phase shifters.  The DQNN and I-DQNN were then leveraged to the imperfect CSI setup by converting the complicated average WSR problem into a normal WSR problem with imperfect CSI as the networks' input.  Simulation results have shown that the DL-based algorithms have better performance than the existing state-of-the-art BCD-based algorithms with comparable complexity in both perfect CSI and imperfect CSI setups. Especially in the imperfect CSI setup, our proposed DL-based algorithms are more stable and robust.

\appendices
\section{Derivation of \eqref{xu17}}
We can further rewrite formula (\ref{xu14}) as
\begin{equation}   \label{xu22}
\begin{aligned}
{Q_A}\left( x \right) &= \sum\limits_{i = 1}^{B - 1} {{a_i}\left[ {\tanh \left( {{c_i}\left( {x - {{\rho}_i}} \right)} \right) + 1} \right]} \\
&= \sum\limits_{i = 1}^{B - 1} {{a_i}\left[ {\frac{{{e^{{c_i}\left( {x - {{\rho}_i}} \right)}} - {e^{ - {c_i}\left( {x - {{\rho}_i}} \right)}}}}{{{e^{{c_i}\left( {x - {{\rho}_i}} \right)}} + {e^{ - {c_i}\left( {x - {{\rho}_i}} \right)}}}} + 1} \right]} \\
&= \sum\limits_{i = 1}^{B - 1} {2 \cdot {a_i}\left[ {\frac{{{e^{{c_i}\left( {x - {{\rho}_i}} \right)}}}}{{{e^{{c_i}\left( {x - {{\rho}_i}} \right)}} + {e^{ - {c_i}\left( {x - {{\rho}_i}} \right)}}}}} \right]}. 
\end{aligned}
\end{equation} 

Let ${u_i} = {e^{{c_i}\left( {x - {{\rho}_i}} \right)}}$ and ${v_i} = {e^{{c_i}\left( {x - {{\rho}_i}} \right)}} + {e^{ - {c_i}\left( {x - {{\rho}_i}} \right)}}$, we have

\begin{equation}   \label{xu23}
\begin{aligned}
{Q_A}\left( x \right) = \sum\limits_{i = 1}^{B - 1} {2 \cdot {a_i}\left( {\frac{{{u_i}}}{{{v_i}}}} \right)}.
\end{aligned}
\end{equation}

We define ${t_i} = \frac{{{u_i}}}{{{v_i}}}$ and obtain the derivative of ${Q_A}\left( x \right)$
\begin{equation}   \label{xu24}
\begin{aligned}
Q_A^{'} = \sum\limits_{i = 1}^{B - 1} {2 \cdot {a_i} \cdot t_i^{'}}. 
\end{aligned}
\end{equation}

It can be easily to get
\begin{equation}   \label{xu25}
\begin{aligned}
t_i^{'} = {\left( {\frac{{{u_i}}}{{{v_i}}}} \right)^{'}} = \frac{{u_i^{'}{v_i} - {u_i}v_i^{'}}}{{v_i^2}},
\end{aligned}
\end{equation}
where $u_i^{'} = {c_i}{e^{{c_i}\left( {x - {{\rho}_i}} \right)}}$ and $v_i^{'} = {c_i}{e^{{c_i}\left( {x - {{\rho}_i}} \right)}} - {c_i}{e^{ - {c_i}\left( {x - {{\rho}_i}} \right)}}$. 

Thus, we can rewrite (\ref{xu25}) as
\begin{equation}   \label{xu26}
\begin{aligned}
t_i^{'} = \frac{{2 \cdot {c_i}}}{{{{\left[ {{e^{{c_i}\left( {x - {{\rho}_i}} \right)}} + {e^{ - {c_i}\left( {x - {{\rho}_i}} \right)}}} \right]}^2}}}.
\end{aligned}
\end{equation}

Then combine (\ref{xu24}) and (\ref{xu26}), we can get
\begin{equation}   \label{xu27}
\begin{aligned}
Q_A^{'}\left( x \right) = \sum\limits_{i = 1}^{B - 1} {\frac{{4 \cdot {a_i} \cdot {c_i}}}{{{{\left[ {{e^{{c_i}\left( {x - {{\rho}_i}} \right)}} + {e^{ - {c_i}\left( {x - {{\rho}_i}} \right)}}} \right]}^2}}}}. 
\end{aligned}
\end{equation}

To avoid that the large exponent in the denominator of (\ref{xu27}) affects the offline training, we made some adjustments to (\ref{xu27})
 \begin{equation}   \label{xu28}
\left.{f_{cons}} = \sum\limits_{i = 1}^{B - 1} {\frac{{4 \cdot {a_i} \cdot {c_i}}}{{{{\left\{ {{e^{\tanh \left[ {{c_i}\left( {x - {{\rho}_i}} \right)} \right]}} + {e^{\tanh \left[ { - {c_i}\left( {x - {{\rho}_i}} \right)} \right]}}} \right\}}^2}}}}.\right.
\end{equation}

\ifCLASSOPTIONcaptionsoff
  \newpage
\fi

\bibliographystyle{IEEEtran}
\bibliography{IEEEref}
\vspace{-10mm}
\end{document}